\documentclass{article}

\def\giorno{5 November 2009}

\def\a{\alpha}
\def\b{\beta}

\def\ga{\gamma}
\def\de{\delta}   
\def\eps{\varepsilon}
\def\la{\lambda}
\def\s{\sigma}

\def\om{\omega}

\def\vth{\vartheta}
\def\vphi{\varphi}
\def\De{\Delta}
\def\Ga{\Gamma}
\def\La{\Lambda}
\def\phi{\varphi}

\def\A{{\mathcal A}}
\def\G{{\mathcal G}}
\def\E{{\mathcal E}}

\def\L{{\mathcal L}}
\def\T{{\rm T}}
\def\G{{\mathcal G}}

\def\X{{\mathcal X}}

\def\pa{\partial}

\def\d{{\rm d}}       

\def\w{\wedge}
\def\ss{\subset}
\def\sse{\subseteq}

\def\({\left(}
\def\){\right)}
\def\[{\left[}
\def\]{\right]}
\def\~#1{\widetilde #1}
\def\.#1{\dot #1}
\def\^#1{\widehat #1}
\def\wt#1{\widetilde #1}

\def\interno{\hskip 2pt \vbox{\hbox{\vbox to .18
truecm{\vfill\hbox to .25 truecm
{\hfill\hfill}\vfill}\vrule}\hrule}\hskip 2 pt}

\def\mapright#1{\smash{\mathop{\longrightarrow}\limits^{#1}}}
\def\mapdown#1{\Big\downarrow\rlap{$\vcenter{\hbox{$\scriptstyle#1$}}$}}

\def\mapne#1{\smash{\mathop{\nearrow}\limits^{#1}}}
\def\mapnw#1{\smash{\mathop{\nwarrow}\limits^{#1}}}
\def\mapse#1{\smash{\mathop{\searrow}\limits^{#1}}}
\def\mapsw#1{\smash{\mathop{\swarrow}\limits^{#1}}}

\def\beq{\begin{equation}}
\def\eeq{\end{equation}}

\def\={\, =\, }

\def\eqref#1{(\ref{#1})}

\begin{document}

\title{Twisted symmetries of differential equations}

\author{Giuseppe Gaeta\\ {\it Dipartimento di Matematica,
Universit\`a degli Studi di Milano,} \\ {\it via Saldini 50, 20133
Milano (Italy)} \\ {\tt giuseppe.gaeta@unimi.it}}

\date{\giorno}

\maketitle

\noindent {\bf Summary.} We review the basic ideas lying at the
foundation of the recently developed theory of twisted symmetries
of differential equations, and some of its developments.

\section*{Introduction}

The study of nonlinear differential equations was the main
motivation to Sophus Lie when he created what is nowadays known as
the theory of Lie groups and Lie algebras, which is by now
recognized as one of the most effective tools for the study of
differential equations -- both in geometrical sense and for what
concerns the search for explicit solutions
\cite{Gbook,KrV,Olv1,Ste}.

The attention of Lie was mainly focused on what we now call {\it
Lie-point symmetries}, but very soon generalizations of these were
investigated, starting with {\it contact symmetries}. The
effectiveness of symmetry methods for differential equations --
greatly increased in recent years by the possibility to perform
the often involved computations required by these via symbolic
manipulation languages, i.e. by computer -- led various authors to
consider generalizations in several directions
\cite{CicK,Cicrev,CRC}; all of these however -- at the exception
of one, i.e. the subject of this paper -- are based on the same
scheme: one considers a vector field $X$ acting on the full phase
manifold $M$  -- the manifold of independent and dependent
variables\footnote{In most symmetry applications, this is just a
linear space $M = B \times U$ with $B$ and $U$ the spaces of
independent and dependent variables respectively. However
equations can be defined on a manifold, in which case $B$ is not a
linear space, or dependent variables take value in a nonlinear
manifold $U$, so that $M = B \times U$ is also not such; or the
equation can be complemented with side conditions, in which case
$M$ is in general a bundle over $B$.} -- and then takes into
account its action on the jet manifold $J^k M$ of suitable order
(i.e. with $k$ the order of the differential equation one wishes
to study), which is obtained by the standard {\it prolongation}
procedure. That is, once we know how $X$ acts on the independent
variables $x^i$ and on the dependent ones $u^a$, we also know how
it acts on the derivatives (of any order) of the $u$'s with
respect to the $x$'s. In computational terms, this is readily
described by the usual {\it prolongation formula}
\cite{Gbook,KrV,Olv1,Ste}; in geometrical terms, this makes use of
the {\it contact structure} which naturally equips the jet
manifold $J^k M$ \cite{ArnGMDE,Olv2}. Thus, the generalizations
lie in the class of admitted vector fields $X$ on $M$, i.e. on
what qualifies (if satisfying certain conditions specific to the
equation under study) as a ``symmetry''; while once this is given,
the action on higher order jet bundles $J^k M$ is just the natural
prolongation of the one in $M$; in other words, we have the vector
field $X^{(k)}$ -- the natural prolongation of $X$ -- acting in
$J^k M$.

Here we are interested in the exception mentioned above, i.e. in
what we would like to call {\bf twisted symmetries} (the reason
for this name will be clear in the following). These were first
proposed by M.C. Muriel and J.L. Romero under the name of
``$C^\infty (M^{(1)})$-symmetries'' or ``$C^\infty$-symmetries''
for short, or finally of ``$\la$-symmetries'', in the framework of
ODEs \cite{MuRom1,MuRomVigo,MuRomJLT,MuRomTMP,PuS}, and then
generalized to PDEs under the name of ``$\mu$-symmetries''
\cite{CGM,Gspt,GM}.

It should be stressed that in this case the generalization with
respect to standard Lie symmetries lies not in the definition of
symmetry\footnote{I.e. in the class of acceptable vector fields or
in the equation-related specific conditions to be satisfied by the
prolonged vector field.}, but in the way vector fields are
prolonged from the phase bundle $M$ to jet bundles $J^k M$. In
this, they are deeply different from other proposed
generalizations of Lie symmetries.

Definition and properties of twisted symmetries will be discussed
below, but we like to recall immediately that these allow to
integrate -- by symmetry methods -- equations which are known not
to admit any standard Lie symmetry \cite{MuRom1,MuRom2,MuRom2b};
needless to say, this is a major reason (but not the only one) for
the interest they raised. It also turned out that twisted
symmetries are interesting from the point of view of the geometry
of differential equations. Here we will try to keep some balance
between these two points of view, the (mostly analytic)
applicative and the (mostly geometric) theoretical one.

\bigskip\noindent
{\bf Acknowledgements.} Many colleagues and friends should be
thanked in connection to this work. First of all I would like to
thank C. Muriel who communicated her results -- both the discovery
of $\lambda$-symmetries and the variational version of her theory
-- prior to publication in two conferences of the SPT series, thus
raising my interest in the matter. I would also like to warmly
thank G. Cicogna and P. Morando, who shared my investigation on
twisted symmetries; and D. Catalano, M.E. Fels, F. Magri, G.
Marmo, P.J. Olver, G. Saccomandi and S. Walcher for several
discussion on these themes over the years. Finally, it is a
pleasure to thank the Editors for their invitation to contribute a
paper to this Special Issue.

\section{Standard prolongations and symmetries}
\label{sec:standardprol}

In this section we will very briefly recall the standard (and well
known) notions of prolongation and symmetry, to the aim of fixing
notation.

We suppose the reader has some familiarity with symmetry of
differential equations and hence do not enter in details; these
can be obtained e.g. from \cite{Gbook,KrV,Olv1,Ste}. Summation
over repeated indices will always be assumed unless otherwise
explicitly stated.

\subsection{Prolongations}

As already stated, we will denote the phase bundle as $M = B
\times U$, where $B$ is the manifold to which the independent
variables belong, equipped with local coordinates $x^i$ ($i =
1,...,q$); and $U$ is the manifold in which the dependent
variables (or fields) take values, which is equipped with local
coordinates $u^a$ ($a=1,...p$).

We consider a (Lie-point) vector field $X$ in $M$; this will be
written in coordinates as \beq\label{eq:vf} X \ = \ \xi^i (x,u) \,
\frac{\pa}{\pa x^i} \ + \ \phi^a (x,u) \, \frac{\pa}{\pa u^a} \ .
\eeq

As well known, given a section $\s_0 = (x, u=f_0(x) ) $ of the
bundle $M=B\times U$, this is transformed under the infinitesimal
action $e^{\eps X}$ of $X$ into the section $\s_\eps = (x,
u=f_\eps(x) ) $ with $$ f_\eps (x) \ = \ f_0 (x) \ + \ \eps \ (
\phi  - \xi^i \, u^a_i )_{0} \ , $$ where the subscript ``$0$''
reminds that the functions $\phi$ and $\xi$ -- as well as the
derivatives $u^a_i$ -- should be computed on $\s_0$, i.e. $$
\phi^a \vert_{\s_0} = \phi^a (x,f_0 (x)) \ , \ \xi^i \vert_{\s_0}
= \xi^i (x,f_0 (x)) \ , \ u^a_i \vert_{\s_0}  = (\pa f_0 (x) / \pa
x) \ .
$$ This is also said by considering the {\it evolutionary
representative} (or vertical representative) $X_v$ of $X$,
$$ X_v \ = \ Q^a \, \frac{\pa}{\pa u^a} \ , \ \ \mathrm{with} \
Q^a \ = \ \phi^a (x,u) \, - \, u^a_i \, \xi^i (x,u) \ . $$

The jet bundles $J^k M$ associated to $M$ can be seen also as
bundles over $B$, $J^k M = B \times U^{(k)}$ (here $U^{(k)}$ is
the manifold of fields and their derivatives of order up to $k$;
we will denote by $U^{[k]}$ the manifold of field derivatives of
order exactly $k$). They can also be seen as bundles over the jet
bundles of lower order, $J^k M = J^{k-1} M \times U^{[k]}$, and it
is this structure which comes into play when considering the
prolongation operation in recursive terms.

The first jet bundle $J M$ is naturally equipped with local
coordinates $(x^i, u^a, u^a_i)$. If $X$ is given in coordinates by
\eqref{eq:vf}, its first prolongation is \beq\label{eq:1prol}
X^{(1)} \ = \ \xi^i (x,u) \, \frac{\pa}{\pa x^i} \ + \ \phi^a
(x,u) \, \frac{\pa}{\pa u^a} \ + \ \psi^a_i (x,u) \,
\frac{\pa}{\pa u^a_i} \ , \eeq where the coefficients $\psi^a_i$
are given by (see below for the derivation of this formula) \beq
\psi^a_i \ = \ D_i \phi^a \ - \ u^a_j \, D_i \xi^j \ . \eeq

Similarly, the jet bundles $J^k M$ are equipped with local
coordinates $(x^i,u^a,u^a_J)$ with $J$ multi-indices of order up
to $k$. If $X$ is given in coordinates by \eqref{eq:vf}, its
$k$-th prolongation is \beq\label{eq:prolvf} X^{(k)} \ = \ \xi^i
(x,u) \, \frac{\pa}{\pa x^i} \ + \ \phi^a (x,u) \, \frac{\pa}{\pa
u^a} \ + \  \sum_{|J|=1}^k \ \psi^a_J (x,u) \, \frac{\pa}{\pa
u^a_J} \ , \eeq where the sum is over all multi-indices $J$ of
order $1 \le |J| \le k$, and the coefficients $\psi^a_J$ are given
by the recursive prolongation formula (here one should understand
$\psi^a_0 \equiv \phi^a$): \beq \label{eq:prol} \psi^a_{J,i} \ = \
D_i \psi^a_J \ - \ u^a_{J,k} \, D_i \xi^k \ . \eeq

Note that, as well known, for vector fields in evolutionary
(vertical) form, the prolongation formula reduces to $$
\psi^a_{J,i} \ = \ D_i \psi^a_J \ . $$

In the case of ODEs, when we have only one independent variable
$x$, the above notation is a bit too heavy, so it is convenient to
also have a simpler notation for this case. We will denote the
derivative of order $j$ as $$ u^a_j \ := \ (\pa^j u^a / \pa x^j)
 \ , $$ and the $k$-th prolongation of a vector field $ X = \xi (x,u)
(\pa / \pa x) + \phi^a (x,u) (\pa / \pa u^a)$ will be written as
\beq X^{(k)} \ = \ \xi (x,u) \, \frac{\pa}{\pa x} \ + \ \phi^a
(x,u) \, \frac{\pa}{\pa u^a} \ + \ \sum_{j=1}^k \ \psi^a_j (x,u)
\, \frac{\pa}{\pa u^a_j}  \eeq with the $\psi^a_j$ satisfying \beq
\label{eq:prolx} \psi^a_{j+1} \ = \ D_x \psi^a_j \ - \ u^a_{j+1}
\, D_x \xi \ . \eeq

Last but not least, we recall that given any two vector fields in
$M$, the commutator of their prolongations is the prolongation of
their commutator, i.e. \beq \label{eq:comm}
\[ X^{(k)} , Y^{(k)} \] \ = \ \( [X,Y] \)^{(k)} \ . \eeq

\subsubsection*{The prolongation formula -- analytic derivation}

We would like to recall how the prolongation formula is obtained
in analytical terms; in a later subsection we will discuss its
geometrical meaning and correspondingly a geometrical derivation.

Let us consider a section $\s = (x,u)$, corresponding locally to a
function $u = f(x)$; let $x_1$ and $x_2 = x_1 + \de x^i$ be two
nearby points in $B$, with $\de x^i$ an infinitesimal displacement
in the direction $x^i$. The corresponding points on $\s$ will be
$p_1 = (x_1, u_1=f(x))$ and $p_2 = (x_2,u_2=f(x_2))$; the partial
derivatives of $u$ on $\s$ are given by
$$ u^a_i (x) \ = \ \lim_{\de x^i \to 0} \, \frac{u^a_2 - u^a_1}{x_2 - x_1}
 \ := \ \lim_{\de x^i \to 0} \, R \ . $$
Needless to say, $u^a_2 = u^a_1 + u^a_i \de x^i + o (\de x^i)$.

If now we act in $M$ by $X$, i.e. by its infinitesimal action
$e^{\eps X}$, the points $p_1$ and $p_2$ are mapped to new points
$\^p_1 = (\^x_1,\^u_1)$ and $\^p_2 = (\^x_2,\^u_2)$ with $$ \^x_j
= x_j + \eps \xi^i (x_j,u_j) \ , \ \ \^u_j = u_j + \eps \phi
(x_j,u_j) \ . $$ We have now to compute (the limit of)
\begin{eqnarray*} \^R = \frac{\^u_2 - \^u_1}{\^x_2 - \^x_1} &=&
\frac{[u_2 + \eps \phi (x_2,u_2)] - [u_1 + \eps (x_1,u_1)]}{[x_2 +
\eps \xi (x_1,u_1)] - [x_1 + \eps (x_1,u_1)]} \\ &=&
\frac{(u_2-u_1) \, + \, \eps [\phi (x_2,u_2) - \phi (x_1,u_1)]
}{(x_2-x_1) \, + \, \eps [\xi(x_2,u_2) - \xi (x_1,u_1)]} \ .
\end{eqnarray*} It results immediately, with all derivatives evaluated in
$(x_1,y_1)$,
\begin{eqnarray*} \xi^k (x_2,u_2) &=& \xi^k (x_1,u_1) + \frac{\pa \xi^k}{\pa x^i}
\de x^i + \frac{\pa \xi^k}{\pa u^b} u^b_i \de x^i = \xi (x_1,u_1)
\ + \ [D_i (\xi)] \, \de x^i \ ; \\ \phi^a (x_2,u_2) &=& \phi^a
(x_1,u_1) + \frac{\pa \phi^a}{\pa x^i} \de x^i + \frac{\pa
\phi^a}{\pa u^b} u^b_i \de x^i = \phi (x_1,u_1) \ + \ [D_i (\phi)]
\, \de x^i \ . \end{eqnarray*} Here and below, $D_i$ denotes the
total derivative with respect to $x^i$, \beq\label{eq:totder} D_i
\ = \ \frac{\pa}{\pa x^i} \ + \ u^a_i \, \frac{\pa}{\pa u^a} \ + \
u^a_{ij} \, \frac{\pa}{\pa u^a_j} \ + \ ... \ . \eeq

We have therefore, recalling $(x_2-x_1) = \de x^i$, writing
$(u_2-u_1) = \de u$ for short, and omitting terms of higher order
in $\eps$,
\begin{eqnarray*} \^R &=& \frac{\de u + \eps
 [(D_i \phi) \de x^i]}{(x_2-x_1) + \eps [(D_i \xi)
\de x^i]} \ = \ \frac{\de u + \eps
 [(D_i \phi) \de x^i]}{\de x^i \ [ 1 + \eps (D_i \xi )]} \\ &=& \frac{[\de u + \eps
 [(D_i \phi) \de x^i] \ [1 - \eps (D_i \xi)]}{\de x^i} \ = \
 \frac{\de u}{\de x^i} \ + \ \eps \[ (D_i \phi) \, - \, \frac{\de
 u}{\de x^i} \, (D_i \xi) \] \\ &=& \ R \ + \ \eps \[ (D_i \phi) \, - \, \frac{\de
 u}{\de x^i} \, (D_i \xi) \] \ . \end{eqnarray*}
This yields, upon taking the limit $\de x^i \to 0$, precisely the
prolongation formula: if $X$ is given in coordinates by
\eqref{eq:vf}, its first prolongation is \beq\label{eq:1prolb}
X^{(1)} \ = \ \xi^i (x,u) \, \frac{\pa}{\pa x^i} \ + \ \phi^a
(x,u) \, \frac{\pa}{\pa u^a} \ + \ \psi^a_i (x,u) \,
\frac{\pa}{\pa u^a_i} \ , \eeq where the coefficient $\psi^a_i$ is
given by \beq \psi^a_i \ = \ D_i \phi^a \ - \ u^a_j \, D_i \xi^j \
. \eeq

The computation for higher order prolongation is exactly the same
(with the role of $u^a$ taken by $u^a_J$).

\subsection{Symmetries}

As well known, a differential equation $E$ of order $k$ defines a
submanifold $S_E$ (of codimension one) in $J^k M$, also called the
{\it solution manifold} for $E$. The same applies for a system
$\De = \{ E^1,...,E^p \}$ of equations; in this case $S_\De$ is of
codimension $p$ (if the system is non degenerate).

If the vector field $X$ on $M$ is such that its $k$-th
prolongation $X^{(k)}$ leaves $S_\De$ invariant, i.e. \beq
\label{eq:symm} X^{(k)} : S_\De \to \T S_\De \ , \eeq we say that
$X$ is a (Lie-point) {\it symmetry of $\De$}. The set of
symmetries of $\De$ will be denoted as $\X_\De$.\footnote{More
precisely, $X$ is a symmetry generator, and $e^{\a X}$ is a
one-parameter symmetry group for $\Delta$. This slight abuse of
notation is commonplace in the literature, and we will keep to
it.}

It follows immediately from this definition -- and from
\eqref{eq:comm} -- that the symmetries of a given $\De$ are a Lie
algebra (under the commutator): \beq\label{eq:liealg} X \in \X_\De
\ , \ Y \in \X_\De \ \Rightarrow \ [X,Y] \in \X_\De \ . \eeq

In order to check if a given vector field $X$ is a symmetry of
$\De$, one just has to check that $$ \[ X^{(k)} (E^a) \]_{S_\De} \
= \ 0 \ . $$

\subsection{The geometric meaning of prolongation}

The prolongation operation was defined before as the lifting of
the action of a vector field on independent variables and fields,
to the derivatives of the latter with respect to the former. In
this way, it was defined in an analytic way. But, prolongation has
a deep and intrinsic geometric meaning as well, and it is
convenient to focus on this in order to consider, later on,
twisted prolongations.

The first jet bundle $J M$ is naturally equipped with local
coordinates $(x^i, u^a, u^a_i)$; in terms of these, the {\it
contact structure} $\Theta^1$ is described by the {\it contact
forms} \beq \theta^a \ = \ \d u^a \ - \ u^a_i \, \d x^i \ . \eeq

The prolongation $X^{(1)}$ of $X$ is the unique vector field in
$J^1 M$ which coincides with $X$ on $M$ and preserves the contact
structure $\Theta^1$.

Similarly, the jet bundles $J^k M$ are equipped with a contact
structure $\Theta^k$; in terms of the local coordinates
$(x^i,u^a,u^a_J)$, the contact structure $\Theta^k$ is described
by the contact forms \beq \theta^a_J \ = \ \d u^a_J \ - \
u^a_{J,i} \, \d x^i \ , \eeq where $0 \le |J| \le k-1$. The
prolongation $X^{(k)}$ of $X$ is the unique vector field in $J^k
M$ which coincides with $X$ on $M$ and preserves the contact
structure $\Theta^k$.

It is maybe worth defining more precisely, for the benefit of the
reader less used to this geometric language, what is meant by
``preservation of the contact structure''.

Consider a set $\{ \vartheta_1 , ... , \vartheta_r \} $ of
generators for $\Theta^k$, and denote by $\E$ the $C^\infty (J^k
M)$ module generated by these same generators\footnote{I.e. the
set of one-forms obtained as sum of the $\vartheta_i$ with
coefficients which are $C^\infty$ functions on $J^k M$, $\theta =
\sum_{i=1}^r \a^i (x,u^{(k)} ) \vartheta^i $.}. We say that a
vector field $Y$ on $J^k M$ preserves $\Theta^k$ if and only if
\beq\label{eq:contact} \L_Y (\vartheta )  \in  \E \ \ \ \forall
\vartheta \in \E \ ; \eeq here and in the following $\L_Y$ is the
Lie derivative under $Y$.

Note, for later reference, that the condition \eqref{eq:contact}
can be expressed equivalently in terms of conditions involving the
commutator of $Y$ with the total derivative operators $D_i$; in
particular, it is equivalent to either one of \begin{eqnarray}
 & & [D_i , Y ] \interno \vth \ = \ 0 \ \ \ \forall \vth \in \E \
 , \label{eq:ceq1} \\
 & & [D_i , Y ] \ = \ h_i^m \, D_m \ + V \ , \label{eq:ceq2}
\end{eqnarray} with $h_i^m \in C^\infty (J^k M)$ and $V$ a vertical vector field
in $J^k M$ (seen as a bundle over $J^{k-1} M$, see above).

\subsubsection*{The prolongation formula -- geometric derivation}

We have recalled above how the prolongation formula is obtained in
analytic terms; here we want to show for comparison how it is
derived in geometric terms, i.e. from \eqref{eq:contact}. We will
again limit to consider the first prolongation, computations being
the same for higher ones.

First of all we note that, from the general properties of the Lie
derivative, we have just to show that $\L_Y (\theta ) \in \E$ for
all $\theta$ in the generating set of the contact structure. In
facts, for $\vartheta = \b_a \theta^a$ with $\b_a \in C^\infty (J
M)$, we have $$ \L_Y (\b_a \theta^a) \ = \ [ \L_Y (\b_a) ]
\theta^a \ + \ \b_a [\L_Y (\theta^a)] \ ; $$ the first term in the
right hand side is by definition in $\E$, and the second is always
in $\E$ if and only if $\L_Y (\theta^a) \in \E$ for all
$a=1,...,p$.

In our case, $ \d \theta^a = \d u^a_i \w \d x^i$. We will write
\beq\label{eq:2} Y \ = \ \xi^i \frac{\pa}{\pa x^i} \ + \ \phi^a
\frac{\pa}{\pa u^a} \ + \ \psi^a_i \frac{\pa}{\pa u^a_i} \ , \eeq
so that $Y \interno \theta^a = \phi^a - u^a_i \xi^i$ and $\d ( Y
\interno \theta^a ) = \d \phi^a - u^a_i \d \xi^i - \xi^i \d
u^a_i$; moreover $Y \interno \d \theta = \xi^i \d u^a_i - \psi^a_i
\d x^i$. Therefore, using the Cartan formula $ \L_Y (\theta) = Y
\interno \d \theta + \d (Y \interno \theta )$, we get immediately
$$ \L_Y (\theta^a) \ = \ \d \phi^a - u^a_j \d \xi^j - \psi^a_i \d
x^i \ . $$ Using $\d u^a = \theta^a + u^a_i \d x^i$, we have $\d
\phi^a = (\pa \phi^a / \pa u^b) \theta^b + (D_i \phi^a) \d x^i$
and $\d \xi^j = (\pa \xi^j / \pa u^b) \theta^b + (D_i \xi^j) \d
x^i$; thus we can rewrite
$$ \L_Y (\theta^a) \ = \ [(\pa \phi^a / \pa u^b) + (\pa \xi^j / \pa u^b)
] \theta^b \ + \ [(D_i \phi^a - u^a_j D_i \xi^j) - \psi^a_i ] \,
\d x^i \ . $$ The first term on the right hand side is surely in
$\E$, by definition, while the second is either zero or surely not
in $\E$. We conclude that $\L_Y (\theta^a) \in \E$ if and only if
$$ \psi^a_i \ = \ (D_i \phi^a) \ - \ u^a_j \ (D_i \xi^j) \ . $$
We have thus obtained again the prolongation formula
\eqref{eq:1prol}.

\section{Lambda-prolongations and symmetries}
\label{sec:Laprol}

The generalization of classical Lie-point symmetries we wish to
consider, i.e. twisted symmetries, modifies the standard
prolongation operation. That is, one considers vector fields $Y$
in $J^n M$ which are not the prolongation of some vector field $X$
in $M$, i.e. $Y \not= X^{(n)}$ for any $X$, yet are related to
such a vector field in a precise manner to be discussed below.
Quite surprisingly, when these are symmetries of a differential
equation $\De$ -- i.e. when $Y : S_\De \to \T S_\De$ -- they are
still effective in obtaining symmetry reductions (for ODEs) or
invariant solutions (for PDEs) of differential equations.

We will start by discussing the first class of twisted symmetries
to be discovered, i.e. $\lambda$-symmetries.

\subsection{The work of Muriel and Romero}

In 2001, C. Muriel and J.L. Romero \cite{MuRom1}, analyzing the
case where $\De$ is a scalar ODE,  noticed a rather puzzling fact.

They substitute the standard prolongation formula \eqref{eq:prolx}
with a ``lambda-prolongation'' formula \beq\label{eq:laprol}
\Psi_{k+1} \ = \ (D_x + \la ) \, \Psi_k \ - \ u_{k+1} \, ( D_x +
\la ) \, \xi \ ; \eeq here $\la$ is a real $C^\infty$ function
defined on $J^1 M$ (or on $J^k M$ if one is ready to deal with
generalized vector fields). For $\la \equiv 0$ one recovers
standard prolongations.

We say that $X$ is a ``lambda-symmetry'' of $\De$ if its
``lambda-prolongation'' $Y$ is tangent to the solution manifold
$S_\De \ss J^n M$.

Then, as mentioned above, it turns out that ``lambda-symmetries''
are as good as standard symmetries for what concerns symmetry
reduction of the differential equation $\De$ and hence
determination of its explicit solutions. As pointed out by Muriel
and Romero, it is quite possible to have equations which have no
standard symmetries, but possess lambda-symmetries and can
therefore be integrated by means of their approach; see their
works \cite{MuRom1,MuRom2} for examples.

It is quite remarkable that the $\la$-symmetries approach is able
to explain the -- rather puzzling -- fact that there were
equations explicitly integrable by quadratures and not possessing
any (standard) symmetry \cite{ASGL,GonGon,GonLop,Olv1}.

\subsection{The invariants-by-differentiation property}

The possibility of using $\la$-symmetries as effectively as
standard ones for order reduction of differential equations has
its origin in the fact the recursion formula allowing to build
higher order differential invariants from lower order ones
\cite{Olv1,Olv2}, also called ``invariants-by-differentiation'',
applies for $\la$-symmetries as well. This point of view was
stressed by Muriel and Romero in another paper \cite{MuRomTMP},
where other examples are also provided.

If $\eta$ and $\zeta$ are differential invariants for the vector
field $Y$, the invariants-by-differentiation property states that
$\rho := (D_x \zeta)/(D_x \eta)$ (with $D_x \eta \not= 0$) will
also be a differential invariant. The proof of this property goes
as follows: acting with $Y$ on $\rho$ we have
$$ Y \( \frac{D_x \zeta}{D_x \eta} \) \ = \ \frac{[Y (D_x \zeta)] (D_x
\eta) - (D_x \zeta) [Y (D_x \eta) ] }{(D_x \eta )^2 } \ . $$ Thus
$\rho$ is a differential invariant if and only if $$ [Y (D_x
\zeta)] (D_x \eta) \ = \  (D_x \zeta) [Y (D_x \eta) ] \ ;  $$ we
can rewrite this equation as $$ (D_x \eta) \( [Y,D_x] (\zeta ) \)
- \( D_x ( Y \zeta) \)  (D_x \eta) \ = \ (D_x \zeta)  \( [Y ,D_x]
(\eta) \) - (D_x \zeta) \( D_x (Y \eta) \)  \ . $$ We assumed
$\eta$ and $\zeta$ are differential invariants for $Y$, hence $Y
(\eta) = Y (\zeta) = 0$, and the above equation reduces to \beq
(D_x \eta) \ \( [Y,D_x] (\zeta ) \)  \ = \ (D_x \zeta) \ \( [Y
,D_x] (\eta) \)  \ . \eeq If $Y$ is a $\la$-prolongation, it
satisfies (up to a term which vanishes on the contact
distribution) \beq\label{eq:lambdacommMR} [D_x,Y] \ = \ \la Y \ +
\ h D_x \eeq with $h$ a smooth function; with this, and recalling
again $Y (\eta) = Y (\zeta) = 0$, the above equation reduces to
the trivial identity
$$ h (D_x \eta) (D_x \zeta ) \ = \ (D_x \zeta) h (D_x \eta) \ . $$

\subsection{Systems of ODE}
\label{sec:muvigo}

Muriel and Romero also considered how the concept of
$\la$-symmetry would extend to systems of ODEs \cite{MuRomVigo}.

They considered a $\la$-prolongation given by
\beq\label{eq:laprol_syst} \psi^a_{k+1} \ = \ (D_x + \la) \psi^a_k
\ - \ u^a_{k+1} \, (D_x + \la) \, \xi \ ,  \eeq i.e. the immediate
generalization of \eqref{eq:laprol}, and studied a system $\De$ of
first order ODEs (which we will also call a dynamical system),
$$u^a_x \ = \ F^a (x,u^1,...,u^r) \ , \ \ \ (a=1,...,r) \ . $$ In
this case, if $X$ is a $\la$-symmetry of the system $\De$, there
exists a change of variables $(y,w) = \b (x,u)$ under which the
system reduces to a system of $(r-1)$ first order ODEs, $$ w^a_y =
G^a (x,w^1,...,w^{r-1} ) \ , \ \ \ (a=1,...,r-1) $$ and the
variable $w^r$ must satisfy an auxiliary ODE $$ H (y,w^r,w^r_y) =
0 \ . $$

\subsection{Lambda symmetries and nonlocal standard symmetries}

In their seminal work, Muriel and Romero also remarked (see sect.5
of  \cite{MuRom1}) that there is an intriguing relation between
$\la$-symmetries and exponential symmetries of differential
equations.

Exponential symmetries \cite{Olv1} represent a specific form of
nonlocal vector fields, \beq\label{eq:nonloc} X \ = \ e^{\int
P(x,u) \, \d x} \ \( \xi (x,u) \frac{\pa }{\pa x} \ + \ \frac{\pa
}{\pa u} \) \ := \ e^{\int P(x,u) \, \d x} \ X_0 \ . \eeq Needless
to say, their prolongation $X^*$ must satisfy the usual symmetry
condition $X^* : S_\De \to \T S_\de$.

Muriel and Romero proved that if $X$ given by \eqref{eq:nonloc} is
an exponential symmetry for $\De$, then $X_0$ defined in
\eqref{eq:nonloc} is a $\la$-symmetry for $\de$, with $$ \la \ = \
P (x,u) \ . $$

As nonlocal symmetries can be used, pretty much as standard
symmetries, to reduce and integrate differential equations
\cite{AM2}, this shows that all the reduction/integration methods
based on nonlocal symmetries (of exponential type)
\cite{AS4,AM1,AM2,Olv1} can be automatically formulated in terms
of $\la$-symmetries.

In a later work \cite{MuRom2007}, Muriel and Romero went further
in their study; in particular they studied the connection between
$\la$-symmetries and the so called ``type I hidden symmetries''
(see below). The key property of use here, shown in their work, is
that if an equation $\d^n u / \d x^n = f (x,u^{(n-1)} ) $ admits a
$\la$-symmetry $X$, then when the equation is written in new
variables $(y,v)$ as $$ \d^n v / \d y^n \ = \ g (y,v^{(n-1)} ) \ ,
$$ the vector field $X$ is a $\^\la$-symmetry for the equation,
with
$$ \^\la \ = \ \frac{\la}{D_x y} \ . $$

Muriel and Romero focused in particular on first order equations;
in this case a theorem by Adam and Mahomed \cite{AM1} gives a
criterion which must be satisfied by a nonlocal symmetry of a
first-order equation in order that a symmetry-related (computable)
transformation maps the latter into an integrable equation; this
theorem can be reformulated (in a simpler way) in the language of
$\la$-symmetries \cite{MuRom2007}. This result was then applied to
relevant equations such as Riccati equations and Abel equations of
the second kind \cite{MuRom2007}.

The relation between $\la$-symmetries and nonlocal standard
symmetries was also investigated by Catalano-Ferraioli
\cite{CF2007} by means of the theory of {\it coverings}
\cite{KrV,Vin1,Vin2}. In this, one embeds $J^k M$ into a space of
higher dimension by adding one or more extra variables $w^\a$ with
their (first) derivatives, and augments the equation under study
with extra equations $$ \d w^\a / \d x \ = \ h^\a (x,u^{(n)},w) \
; $$ nonlocal symmetries can then be expressed as local symmetries
of the augmented system. The connection with $\la$-symmetries
arises (with a single extra variable $w$) when one chooses $h =
\la$. We refer to the original paper \cite{CF2007} for details.

\subsection{Lambda symmetries and other types of symmetries}

Lambda symmetries turned out to be relevant, and providing a way
to soundly understand, also other kinds of symmetries. Here we
will just briefly mention these interrelations, referring the
reader to the original papers.

Gandarias, Medina and Muriel \cite{GMM} discussed the relation
between $\la$-symmetries and the so called {\it potential
symmetries} of differential equations \cite{BR}. They also
discussed how this relation can be of help in the integration of
differential equations not possessing Lie-point symmetries. The
same problem was also tackled in a paper by Muriel and Romero
\cite{MuRomJLT}, where they also discuss the relation with
potential symmetries.

Similarly, $\la$-symmetries are instrumental in integrating
equations with non-solvable symmetry algebras; this point was
discussed by Muriel and Romero both in general terms \cite{MuRom2}
and for specific algebras relevant in Physics (and in other
applications as well) \cite{MuRomSPT,MuRom2b}.

Lambda symmetries also bear some interesting relation to so called
{\it hidden symmetries} (such as symmetries which got lost in the
reduction process) \cite{AS}. See \cite{AS2,AS3} for a discussion.

Hidden symmetries also have relations \cite{HA} with so called
{\it solvable structures} \cite{BH,BP}; these in turn have been
very recently studied in connection to $\la$-symmetries
\cite{CFM}. We refer the reader to the original paper \cite{CFM}
for this matter.

\subsection{Lambda symmetries and integrating factors}

It turns out that $\la$-symmetries are also relevant for the
determination of {\it integrating factors} for higher order ODEs.
As well known, any first order ODE $$ N(x,u) \ \frac{\d u}{\d x} \
+ \ M(x,u) \ = \ 0 $$ admits an integrating factor, i.e. a scalar
function $\rho (x,u)$ such that $$ \rho \ \[ N (x,u) \, \d u \ + \
M(x,u) \, \d x \] \ = \ \d F $$ for some function $F = F(x,u)$.

The integrating factor of an $n$-th order equation $$
N(x,u,u',...,u^{(n-1)} ) \, \frac{\d^n u}{\d x^n} \ + \
M(x,u,u',...,u^{(n-1)} ) \ = \ 0 $$ is a scalar function $\rho$
such that
$$ \mathcal{D}_u \ \( \rho ( M + N u_n ) \) \ = \ 0 \ , $$
where $\mathcal{D}_u $ is the {\it variational derivative},
$$ \mathcal{D}_u \ = \ \frac{\de}{\de u} \ = \ \frac{\pa}{\pa u}
\ - \ D_x \, \frac{\pa }{\pa u_x} \ + \ D_x^2 \, \frac{\pa }{\pa
u_{xx} } \ + \ ... \ .
$$

The problem of integrating factors for higher order
equations\footnote{For the relation between ordinary symmetries
and integrating factors, see e.g. \cite{LB}.} was tackled by
Muriel and Romero using symmetries and $\la$-symmetries
\cite{MuRomIF}.

In particular, they were able to identify, given an integrating
factor, a $\la$-symmetry associated to it; and conversely, given a
$\la$-symmetry, a corresponding integrating factor. As any ODE
admits $\la$-symmetries, this implies in particular that --
similarly to what happens for first order ODEs, as stated by the
very classical result by Clairaut -- any differential equation of
order $n$, $$ \frac{\d^n u}{\d x^n} \ = \ f (x,u^{(n-1)} ) $$ with
$f$ analytic in some open subset of $J^{n-1} M$, admits an
integrating factor $\rho (x,u^{(k)})$ with some $k < n$.
\footnote{After the first version of this paper was submitted,
Muriel and Romero published another paper, in which they study in
depth the interrelations between $\la$-symmetries, integrating
factors and first integrals of second order equations
\cite{MuRom2009}; this also includes application to Ermakov-Pinney
equation. Here we can just urge the reader to read this paper.}

\subsection{Lambda symmetries and first integrals}

As mentioned above, $\la$-symmetries can also be considered for
systems of ODEs, and in particular for dynamical systems. In this
case, one is specially interested in first integrals. The relation
between these and $\la$-symmetries was considered by Zhang and Li
\cite{ZL}.

A dynamical system \beq\label{eq:ZL} u^a_x \ = \ f^a (x,u) \ ;
\eeq this is represented by the dynamical vector field $F =
f^a(x,u) (\pa / \pa u^a)$. With $\L$ the Lie derivative, a vector
field $Z = \phi^a (x,u) (\pa / \pa u^a) $ is a $\la$-Liouville
vector field (or simply a Liouville vector field when $\la = 0$)
if
$$ \( \frac{\pa}{\pa x} + \L_F + \la \) \, Z \ + \ (\mathrm{div} F)  \,
Z \ = \ 0 \ . $$

If $X$ is a $\la$-symmetry for \eqref{eq:ZL}, then $( \pa_x + \L_F
+ \la ) X = 0$. Moreover, let $Z$ be Liouville for \eqref{eq:ZL}
and the scalar function $p(x,u)$ satisfy the equation $ (\pa p /
\pa x) + \mathrm{div} (p F) = 0$; then the vector field $X :=
(1/p) Z$ is a $\la$-symmetry for \eqref{eq:ZL} \cite{ZL}.

Now, let $\ga$ be a scalar analytic function, solution to $$ (
\pa_x + \L_F ) \ga = \la \ga \ . $$ Under the assumption that $\ga
Z$ is divergence-free, Zhang and Li propose a way, starting from a
Liouville vector field $Z$ and $\ga$, and having determined the
$n-1$ invariants of the vector field $Z$ (obtained by solving the
associate characteristic equation), to build $(n-1)$ first
integrals for \eqref{eq:ZL}. Note that this associates to a single
symmetry not one, but $(n-1)$ first integrals; on the other hand,
the requirement of determining a $\la$-Liouville vector field {\it
and} its characteristic lines is a rather strong one; in
particular, solving for the invariants of $Z$ is in general not
any easier than integrating \eqref{eq:ZL}.

\section{The geometric meaning of lambda symmetries}
\label{sec:geomlambda}

The results reviewed in the previous section are rather
impressive, in particular in that they were able to include in the
symmetry theory of differential equations some features which
appeared, beforehand, to be definitely out of its reach.

Nevertheless, these results were reached by an essentially
analytic approach, and thus shed little light on the geometric
meaning of lambda-prolongations and symmetries. Also, they were
limited to ODEs; one would hope that understanding the geometry of
lambda-prolongation would allow for their extension to PDEs. In
this respect, two papers were quite instrumental in bridging the
gap between analysis and geometry of $\la$-prolongations, and
between ODEs and PDEs as far as $\la$-symmetries were concerned.

\subsection{The work of Pucci and Saccomandi}

In 2002, Pucci and Saccomandi \cite{PuS} devoted further study to
lambda-symmetries, and stressed a very interesting geometrical
property of lambda-prolongations: that is, lambda-prolonged vector
fields in $J^n M$ can be characterized as the {\it most general}
vector fields in $J^n M$ which have the same characteristics as
some standardly-prolonged vector field.

We stress that if $Y$ is the lambda-prolongation of a vector field
$X$ in $M$, then the characteristics of $Y$ will not be the same
as those of the standard prolongation $X^{(n)}$ of $X$, but as
those of the standard prolongation $\wt{X}^{(n)}$ of a generally
different (for $\la$ nontrivial) vector field $\wt{X}$ in $M$.

This property can also be understood by recalling \eqref{eq:prol},
\eqref{eq:prolx} and making use of a general property of Lie
derivatives: indeed, for $\a$ any form on $J^n M$, \beq \L_{\la Y}
(\a) \ = \ \la Y \interno \d \a \, + \, \d (\la Y \interno \a ) \
= \ \la \, \L_Y (\a ) \, + \, \d \la \w (Y \interno \a ) \ . \eeq

\subsection{The work of Morando}

It was noted \cite{GM,MuRomVigo} that lambda-prolongations can be
given a characterization similar to the one discussed above for
standard prolongations; that is, with $h_i^m$ and $V$ as above,
\eqref{eq:laprol} is equivalent to either one of\footnote{Note
that the second of these relations was already remarked -- and
used -- by Muriel and Romero; see eq.\eqref{eq:lambdacommMR}
above.}
\begin{eqnarray*}
 & & [D_x , Y ] \interno \vth = \la (Y \interno \vth ) \ \ \ \forall \vth
\in \E \ , \\ & & [D_x , Y ] = \la  Y +  h_i^m D_m + V  \ .
\end{eqnarray*}

This, as remarked by Morando, also allows to provide a
characterization of lambda-prolonged vector fields in terms of
their action on the contact forms, analogously to
\eqref{eq:contact}.

In this context, it is natural to focus on the one-form $ \mu :=
\la \d x $; note this is horizontal for $J^n M$ seen as a bundle
over $B$, and obviously satisfies $D \mu = 0$, with $D$ the total
exterior derivative operator. Then, {\it $Y$ is a lambda-prolonged
vector field if and only if,  for all $\vth \in \E$,
\beq\label{eq:la_mor} \L_Y (\vth ) + (Y \interno \vth ) \mu \in \E
\ .  \eeq}

\section{Twisted symmetries for PDEs: mu-pro\-lon\-ga\-ti\-ons and mu-symmetries}
\label{sec:muprol}

The result given above about the geometrical characterization
\eqref{eq:la_mor} of $\la$-symmetries, immediately opens the way
to extend $\la$-symmetries to PDEs \cite{GM}. As here the main
object will be the one-form $\mu$, we prefer to speak of
``$\mu$-prolongations'' and ``$\mu$-symmetries''.

\subsection{Mu-prolongations}

Consider a semi-basic one-form \beq\label{eq:mu} \mu \ := \ \la_i
\, \d x^i \eeq on $(J^n M, \pi_n , B)$, satisfying $D \mu = 0$.
Then we say that the vector field $Y$ in $J^n M$ {\it
$\mu$-preserves the contact structure} if and only if, for all
$\vth \in \E$, \beq\label{eq:mucont} \L_Y (\vth ) \ + \ (Y
\interno \vth ) \, \mu \ \in \ \E \ . \eeq

Note that $D \mu = 0$ means $D_i \la_j = D_j \la_i $ for all
$i,j$; hence {\it locally} $\mu = D \Phi$ for some smooth real
function $\Phi$.

With standard computations \cite{GM}, one obtains that
\eqref{eq:mucont} implies the {\bf scalar $\mu$-prolongation
formula} \beq\label{eq:muprolscal} \Psi_{J,i} \ = \ (D_i + \la_i )
\, \Psi_J \ - \ u_{J,m} \, ( D_i + \la_i ) \, \xi^m  \ . \eeq

Let $Y$ as in (2) be the $\mu$-prolongation of the Lie-point
vector field $X$ (1), and write the standard prolongation of the
latter as $ X^{(n)} = \xi^i  \pa_i + \Phi_J  \pa_u^J$; note that
$\Psi_0 = \Phi_0 = \phi$. We can obviously always write $\Psi_J =
\Phi_J + F_J$, and $F_0 = 0$. Then it can be proved \cite{GM} that
the difference terms $F_J$ satisfy the recursion relation \beq
F_{J,i} \ = \ (D_i + \la_i ) F_j + \la_i D_J Q \eeq where $Q :=
\phi - u_i \xi^i$ is the characteristic \cite{Gbook,Olv1,Ste} of
the vector field $X$.

This shows at once that {\it the $\mu$-prolongation of $X$
coincides with its standard prolongation on the $X$-invariant
space $I_X$}; indeed, $I_X \ss J^n M$ is the subspace identified
by $D_J Q = 0$ for all $J$ of length $0 \le |J| < n$. It follows
that the standard PDE symmetry reduction method
\cite{Gbook,Olv1,Ste} works equally well when $X$ is a
$\mu$-symmetry of $\De$ as in the case where $X$ is a standard
symmetry of $\De$; see e.g. \cite{CGM,GM} for
examples.\footnote{The concept of $\mu$-symmetries is also
generalized to an analogue of standard conditional and partial
symmetries \cite{CicK,CGpar}, i.e. partial (conditional)
$\mu$-symmetries \cite{CGM}.}

\subsection{Mu-symmetries for systems of PDEs}

The developments described in the previous subsection do not
include the case of (systems of) PDEs for several dependent
variables, i.e. the case with $q>1$ in our present notation. This
was dealt with in two works \cite{CGM,GM}, to which we refer for
details.

In this case the relevant contact forms are \beq \vth^a_J \ := \
\d u^a_J \ - \ u^a_{J,i} \, \d x^i \ , \eeq and it is convenient
to see them as the components of a vector-valued contact form
$\vth_J$ \cite{Str}. We will denote by $\Theta$ the module over
$q$-dimensional smooth matrix functions generated by the $\vth_J$,
i.e. the set of vector-valued forms which can be written as $\eta
= (R_J)^a_b \vth^b_J $ with $R_J : J^n M \to \mathrm{Mat}(q)$
smooth matrix functions.

Correspondingly, the fundamental form $\mu$ will be a horizontal
one-form with values in the Lie algebra $g \ell(q)$ (the algebra
of the group $GL(q)$, consisting of non-singular $q$-dimensional
real matrices) \cite{Str}. We will thus write \beq\label{eq:14}
\mu \ = \ \La_i \, \d x^i \eeq where $\La_i$ are smooth matrix
functions satisfying additional compatibility conditions stated
and discussed below.

We will say that the vector field $Y$ in $J^n M$ $\mu$-preserves
the vector contact structure $\Theta$ if, for all $\vth \in
\Theta$, \beq \L_Y (\vth ) \ + \ \( Y \interno (\La_i)^a_b \vth^b
\) \ \d x^i  \ \in \ \Theta \ ; \eeq this should be compared to
standard preservation of the contract structure in the form
\eqref{eq:contact}.

In terms of the coefficients of $Y$, see \eqref{eq:2}, this is
equivalent to the requirement that the $\Psi^a_J$ obey the {\bf
vector $\mu$-prolongation formula} \beq\label{eq:16} \Psi^a_{J,i}
\ = \ (\nabla_i)^a_b \, \Psi^b_J \ - \ u^b_{J,m} \, [(
\nabla_i)^a_b \, \xi^m ] \ , \eeq where we have introduced the
(matrix) differential operators
$$ \nabla_i \ := \ I \, D_i \ + \ \La_i \ ,  $$
with $I$ the $q \times q$ identity matrix.

We note for later reference that for vertical vector fields $X =
Q^a (\pa / \pa u^a)$, \eqref{eq:16} yields for the coefficients of
the first prolongation $Y = X + \psi^a_i (\pa / \pa u^a_i)$,
simply  \beq\label{eq:vproluno} \psi^a_i \ = \ (\nabla_i)^a_b \,
Q^b \ = \ D_i \, Q^a \ + \ (R_i)^a_{\ b} \, Q^b \ . \eeq

If $Y$ is the $\mu$-prolongation of the vector field $X$, and $Y :
S_\De \to \T S_\De$, we say that $X$ is a $\mu$-symmetry for
$\De$.

If again we consider a vector field $Y$ as in \eqref{eq:2} which
is the $\mu$-prolongation of a Lie-point vector field $X$, and
write the standard prolongation of the latter as $ X^{(n)} = \xi^i
\pa_i + \Phi^a_J  \pa_a^J$ (with $\Psi^a_0 = \Phi^a_0 = \phi^a$),
we can write $\Psi^a_J = \Phi^a_J + F^a_J$, with $F^a_0 = 0$. Then
the difference terms $F_J$ satisfy the recursion relation
\beq\label{eq:F} F^a_{J,i} \ = \ \de^a_b \[ D_i (\Ga^J)^b_c \]
(D_j Q^c ) \, + \, (\La_i)^a_b
\[ (\Ga^J)^b_c (D_J Q^c) + D_J Q^b \] \ , \eeq where $Q^a := \phi^a
- u^a_i \xi^i$, and $\Ga^J$ are certain matrices (see ref.
\cite{GM} for the explicit expression). This, as for the scalar
case, shows that the $\mu$-prolongation of $X$ coincides with its
standard prolongation on the $X$-invariant space $I_X$; hence,
again, the standard PDE symmetry reduction method works equally
well for $\mu$-symmetries as for standard ones. See ref. \cite{GM}
for examples.

\subsection{Compatibility condition, and gauge equivalence}

As mentioned above the form $\mu$, see \eqref{eq:14}, is not
arbitrary: it must satisfy a compatibility condition (this
guarantees the $\Psi^a_J$ defined by \eqref{eq:16} are uniquely
determined), expressed by \beq\label{eq:compcon} \[ \nabla_i ,
\nabla_k \] \ \equiv \ D_i \La_k \, - \, D_k \La_i \ + \ [\La_i ,
\La_k ] \ = \ 0 \ . \eeq

It is quite interesting to remark \cite{CGM} that this is nothing
but the coordinate expression for the horizontal Maurer-Cartan
equation\footnote{This expresses the requirement that the standard
Maurer-Cartan equation is satisfied modulo contact forms, i.e. $\d
\mu + (1/2) [\mu , \mu] \in \E$.} \beq\label{eq:20} D \mu \ + \ {1
\over 2} \ [ \mu , \mu ] \ = \ 0 \ . \eeq

Based on this condition, and on classical results of differential
geometry \cite{CCL,EGH,Sha} and a theorem by Marvan \cite{Mar}, it
follows that locally in any contractible neighborhood $A \sse J^n
M$, there exists $\ga_A : A \to GL(q)$ such that (locally in $A$)
$\mu$ is the Darboux derivative of $\ga_A$.\footnote{In a
different context, it should also be stressed that the
compatibility condition \eqref{eq:compcon} -- or equivalently the
appearance of the horizontal Maurer-Cartan equation -- also hints
at a relation between $\mu$-symmetries and {\it zero curvature
representations } for PDEs \cite{Mar}, and hence the theory of
integrable systems. This aspect of $\mu$-symmetries has not been
studied, and could provide interesting results.}

In other words, any $\mu$-prolonged vector field is {\it locally}
gauge-equivalent to a standard prolonged vector field \cite{CGM},
the gauge group being $GL(q)$.

It should be mentioned that when $J^n M$ is topologically
nontrivial, or $\mu$ presents singular points, one can have
nontrivial $\mu$-symmetries; this is shown by means of very
concrete examples in \cite{CGM}.

Finally, we note that when we consider symmetries of a given
equation $\De$, the compatibility condition \eqref{eq:20} needs to
be satisfied only on $S_\De \sse J^n M$. Indeed when $\mu$ is not
satisfying everywhere \eqref{eq:20}, $\mu$-symmetries can happen
to be gauge-equivalent to standard {\it nonlocal symmetries} of
exponential form, as noted in \cite{MuRom1} and also remarked in
\cite{CGM}; in this respect, see also sect.\ref{sect:nonlocal}
below.

\subsection{The work of Cicogna: rho-symmetries}

The approach of $\mu$-symmetries is inherently
multi-dimensional\footnote{Actually the one-dimensional case is a
very degenerate one, first of all because the compatibility
condition is trivially satisfied.}; one could thus hope that it
would provide better results for the analysis of systems of ODEs.

This task was undertaken by Cicogna, who focused on systems of
first order ODEs \cite{Cicrho}. He considered in particular the
possibility of reducing such a system; reduction is achieved by
passing to suitable (symmetry-adapted) coordinates.

Consider a system of $n$ equations for $u^a (x)$, $a=1,...,n$,
given by \beq\label{eq:cicsyst1} F^a (x,u,u_x) \ = \ 0 \ \ \ (a
=1,...,n) \eeq and which admits a $\mu$-symmetry $X$, with $\mu =
\La \d x$; let $(y , w^1,..., w^{n-1} , z)$ be the symmetry
adapted coordinates, with $y$ the new independent variable and $z$
the invariant dependent variable. Then the system
\eqref{eq:cicsyst1}, when written in the $(y,w,z)$ variables,
depend only on the $(y,w)$ variables and on the $n$ first order
differential invariants $\zeta^a$, obtained solving the
characteristic equation
$$ \d z \ = \ \frac{\d w^a_y}{M^{(a)}} \ = \ \frac{\d z_y}{M^n}
\ . $$ Here the $M^{(a)}$ are matrix functions given by $$ M^{(a)}
\ = \ \frac{\pa w^a}{\pa u^b} (\La Q)^b \ \ (a = 1,...,n-1) \ , \
\ M^{(n)} \ = \ \frac{\pa z}{\pa u^b} (\La Q)^b \ . $$ We have
denoted by $Q$ the characteristic of the vector field $X$: if this
is written in the form \eqref{eq:vf}, we have $Q^a = \phi^a - \xi
u^a_x$.

This results gets more clear if instead of general first order
systems \eqref{eq:cicsyst1} we consider dynamical systems,
\beq\label{eq:cicsyst2} u^a_x \ = \ f^a (x,u) \ . \eeq In this
case one would like that the transformation to symmetry adapted
coordinates preserves the functional form of the system, i.e. to
get
$$ w^a_y = g^a (y,w,z) \ (a=1,...,n-1) \ , \ \ z_y = h (y,w,z) \ .
$$ In this case, assuming again \eqref{eq:cicsyst2} admits $X$ as
a $\mu$-symmetry with $\mu = \La \d x$, we actually have $$ (\pa
g^a / \pa z) = M^{(a)} \ , \ \ (\pa h / \pa z) = M^{(n)} \ . $$ If
$M^{(a)}=0$ for all $a$, so that no $f^a$ depends on $z$, the
system splits into an $(n-1)$-dimensional system for the $w^a$
plus a scalar ODE $z_y = h (y,w,z)$ which can be seen as a
``reconstruction equation'' whose solution allows to pass from
solutions to the reduced system to solutions to the full one and
hence to the original system \eqref{eq:cicsyst2}.

This result is rather similar to what is obtained for standard
symmetries, but it should be noted that while in that case the
``last equation'' reduces to a quadrature\footnote{If $\La=0$, so
that we have a standard symmetry, we obviously have $M^{(a)} =
M^{(n)} = 0$, and the reconstruction equation is a quadrature.
Conversely, one can prove that if $M^{(a)} = M^{(n)} = 0$, then
$X$ is a standard symmetry \cite{Cicrho}.}, in this case we have
in general to deal with a possibly nontrivial equation. When $\La
= \la I$, one is reduced to the situation studied by Muriel and
Romero \cite{MuRomVigo}, see sect.\ref{sec:muvigo}.

As stressed by Cicogna, while in general $\la$-symmetries lead to
a lowering of the order of the equation under study (or of one of
the equations in the system), $\mu$-symmetries for dynamical
system do not reduce the order of the system, but rather split it
into a reduced one and a reconstruction equation. Thus, he
suggests that this special kind of twisted symmetry is given the
name of $\rho$-symmetries, where $\rho$ stands for ``reducing''
(in this context, $\la$ would also mean ``lowering'')
\cite{Cicrho}.

\subsection{Twisted symmetries and nonlocal Lie-point symmetries for PDEs}
\label{sect:nonlocal}

The connection between twisted symmetries and standard nonlocal
ones, already discussed in the ODEs case (see above), is also
present for PDEs.

In this case one considers nonlocal vector fields of the form
$$ X \ = \ e^{\int P_i (x,u^{(n)}) \ d x^i } \ Z $$
with $X_0$ a local vector field in $M$. This is a nonlocal
exponential symmetry of $\De$ if $Z$ is a symmetry of $\De$ and
$P$ satisfies, at least on the solution manifold $S_\De$, the
compatibility condition $D_i P_j = D_j P_i$.

Then, consider the form $\mu = P_i \d x^i$. If $Z$ is a
$\mu$-symmetry of $\De$, and $D_i P_j = D_j P_i$ on $S_\De$, then
$X$ is a (standard) nonlocal exponential symmetry for $\De$; and
conversely, if $X$ as above is a (standard) nonlocal exponential
symmetry for $\De$, then $Z$ is a $\mu$-symmetry for $\De$
\cite{CGM}.

%
%
%
%
\subsection{Twisted symmetries and gauging the exterior
derivative}

It was remarked by Morando \cite{Mor2007} that $\mu$-prolongations
and symmetries can also be described in terms of a deformation of
the derivation operations (Lie derivative and exterior
derivative).

In fact, one can consider a deformed differential based on a
smooth function $f$ and corresponding to a ``gauging'' by $f$:
\beq \d^{\d f} \b \ := \ e^{-f} \ \d \( e^f \, \b \) \ = \ \d \b \
+ \ \d f \w \b \ . \eeq It follows by straightforward computation
that this is a first order differential operator, preserving wedge
product, and satisfying $\d^{\d f} \circ \d^{\d f} = 0$ (so that
one can build complexes and a cohomology on its basis).

Similarly, one can consider a deformed Lie derivative based on a
smooth function $f$, defined on forms as \beq \L_X^{\d f} \, \b \
:= \ e^{-f} \ \L_{(e^f X)} \, \b \ = \ \L_X \, \b \ + \ \d f \w (X
\interno \b ) \ . \eeq This coincides with the ordinary Lie
derivative on functions, and in general satisfies
$$ \begin{array}{l}
\L_X^{\d f} \ (\b_1 + \b_2 ) \ = \ \L_X^{\d f} \, \b_1 \ + \
\L_X^{\d f} \, \b_2 \ ; \\
\L_X^{\d f} \ (\b_1 \w \b_2 ) \ = \ \( \L_X^{\d f} \b_1 \) \, \w
\, \b_2 \ + \ \b_1 \, \w \, \( \L_X^{\d f} \b_2 \) \ . \end{array}
$$
This deformed Lie derivative will be defined on vector fields as
\beq \L_X^{\d f} \, (Y) \ := \ e^{-f} \L_{(e^f X)} \, Y \ = \ \L_X
\, Y \ - \ (Y \interno \d f) \, X \ . \eeq It is immediate to
check \cite{Mor2007} this entails, for all vector fields $X,Y$ and
forms $\b$ on $M$,
$$ \L_X^{\d f} \, (Y \interno \b) \ = \ \L_X^{\d f} \, (Y)
\interno \b \ + \ Y \interno \( \L_X^{\d f} \, \b \) \ . $$

Moreover, Cartan formulas -- with the deformed exterior derivative
taking the place of the ordinary one -- hold for the deformed Lie
derivative: \beq \begin{array}{l} \L_X^{\d f} \, (\b
) \ = \ X \interno \d \b \ + \ \d^{\d f} \, (Y \interno \b) \ , \\
\L_X^{\d f} \, (\d \b) \ = \ \d^{\d f} \ \( \L_X^{\d f} \, (\b) \)
\ . \end{array} \eeq

As stressed by notation, the deformed derivatives actually depend
on $\d f$ rather than on $f$; it is thus quite natural to define
them depending on a general closed (but not necessarily exact)
one-form $\mu \in \Lambda^1 (M)$. This yields
\begin{eqnarray}
 & & \d^\mu \b \ := \ \d \b \ + \ \mu \w \b \ ; \\
 & & \L^\mu_X \, \b \ := \ \L_X \, \b \ + \ \mu \w (X \interno \b)
 \ , \\
 & & \L^\mu_X \, Y \ := \ \L_X \, Y \ - \ (Y \interno \mu) \, X \
 . \end{eqnarray}
One can check, again by explicit computation \cite{Mor2007} that
$\d \mu = 0$, $\d^\mu \circ \d^\mu = 0$, and $\L^\mu_X \d \b =
\d^\mu (\L_X^\mu \b)$ are all equivalent.

Needless to say, these definitions can be extended from $M$ to
$J^k M$, i.e. to Jet bundles of any order, in a rather obvious
way. In this case denote, as earlier on, by $\E$ the contact ideal
(the ideal generated by the contact forms). Then, $\d \mu \in \E$,
$\d^\mu (\d^\mu \b) \in \E$ and $\L_X^\mu (\d \b) - \d^\mu
(\L_X^\mu \b) \in \E$ are all equivalent.

Consider now a vector field $Y$ on $J^n M$, which leaves $M$
invariant and which reduces to $X$ when restricted to $M$. It can
be proven \cite{Mor2007} that $Y$ is the $\mu$-prolongation of $X$
if and only if the deformed Lie derivative preserves $\E$, i.e.
$$ \L_Y^\mu \ (\theta ) \ \in \ \E \ \ \ \forall \theta \in \E \ . $$
Moreover, with $\mathcal{D}$ the distribution generated by the
operators $D_i$, $Y$ is the $\mu$-prolongation of $X$ if and only
if
$$ \L_Y^\mu \ (Z) \ \in \ \mathcal{D} \ \ \ \forall Z \in \mathcal{D} \ .
$$

This formalism is also helpful when one considers variational
twisted symmetries; the latter will be discussed in a later
section.

\section{Twisted symmetries and gauged vector fields}
\label{sec:gaugeA}

\def\eb{{\bf e}}
\def\fb{{\bf f}}

The discussion so far, in particular for ODEs, mentioned at
several points the concept of {\it gauge transformations}. This is
one of the central ideas in $20^{th}$ century Physics
\cite{CCL,Nak,NaS}, and in a way it is surprising that it has been
absent so far in the symmetry theory of differential
equations.\footnote{It is maybe worth warning the experienced
reader that the gauge transformations to be considered here are,
in general, of a more general type than standard Yang-Mills ones.}

As stated above, see sect.\ref{sec:muprol}, $\mu$-prolonged vector
fields are (locally) gauge equivalent to standard-prolonged ones.
More precisely, if $Y$ is the $\mu$-prolongation of a vector field
$X$, then there are vector fields $W$ and $Z$, gauge-equivalent
via the same gauge transformation (acting respectively in $\T
(J^k) M$ and in $\T (M)$) to $Y$ and $X$, and such that $W$ is the
standard prolongation of $Z$. This is schematically summarized in
the following diagram:
$$ \matrix{X & & \mapright{\ga} & & Z \cr
\mapdown{\mu-{\mathrm{prol}}} & & & & \mapdown{{\mathrm{prol}}}
\cr Y & & \mapright{\ga^{(k)}} & & W \cr} \ . $$ For these
considerations, it is convenient to deal with evolutionary
representatives of vector fields \cite{Olv1}, which we will
implicitly do.

We recall that the gauge group $\Ga$ (modelled over a Lie group
$G$) acts in the same way on the vector $\{ \phi^1 , ... , \phi^q
\}$ of the components of the vector field $X$ in $M$, and on the
vectors $\{ \psi^1_J , ... , \psi^q_J \}$ of components (relative
to a given multi-index $J$, i.e. to partial derivatives with
respect to the same array of independent variables) of the vector
field $Y$ in $J^k M$. One also says that $\Ga$ acts via a {\it Jet
representation}.

It is thus quite natural to consider a framework in which gauge
transformations are also taken into account; that is -- similarly
to what happens in gauge theories of Theoretical Physics -- one
introduces, beside dependent and independent variables, also {\it
gauge variables}; these keep track of the reference frame changes
\cite{Ggau1}.\footnote{It has to be stressed that here one
considers changes of reference frame, and {\it not} changes of
variables.}

At difference with the approach of gauge theories, however, the
tradition in Applied Mathematics (in general, and in the symmetry
theory of differential equations in particular) is to deal with
standard partial derivatives rather than with {\it covariant
derivatives}. In the specific prolongations framework, this means
that the total derivatives appearing in the prolongation formula
\eqref{eq:prol} $$ \psi^a_{J,i} \ = \ D_i \psi^a_J $$ (for
evolutionary vector field the prolongation formula reduces to
this) change when we change frame.

Let us choose a representation for $G$, and a basis $\{ L_1 , ...
, L_r \}$ of (left-invariant) vector fields for the Lie algebra
$\G$ of $G$; then the gauge transformations can be written as \beq
\label{eq:gauge} S(x) \ = \ \exp [\a^i (x) L_i ] \ . \eeq

Denote by $\{ \eb_1 , ... , \eb_p \}$ the reference frame for the
tangent space $\T U$ to the manifold of dependent variables, and
by  $\{ \fb_1 , ... , \fb_p \}$ another frame, related to the
previous one by $$ \fb_a (x) \ = \ S_a^{\ b} (x) \ \eb_b (x) \ .
$$ A field $\Phi (x)$ will be given by $$ \Phi (x) \ = \ u^a (x) \
\fb_a (x)
$$ with $u^a$ the components in the $\fb$ frame; in the $\eb$
frame, we would have
$$ \Phi (x) \ = \ u^a (x) \ S_a^{\ b} (x) \ \eb_b (x) \ := \ v^a
(x) \ \eb_a (x) \ . $$ If now we look at $x$-derivatives, we get
(we omit to indicate the $x$ dependence for the sake of notation)
$$ D_i \, \Phi \ = \ u^a_i \ \fb_a $$ in the $\fb$ frame; while
in the $\eb$ frame we get, using the representation
\eqref{eq:gauge}, \beq D_i \, \Phi \ = \ \[ \a^m \ (L_m^T)^a_{\ b}
\ v^b_i \ + \ \a^m_i \ (L_m^T)^a_{\ b} \ v^b \] \ . \eeq That is,
field derivatives do not change in the same way as the field; as
well known this phenomenon can be eliminated by considering {\it
covariant derivatives} instead of standard ones
\cite{CCL,Nak,NaS}.

We will thus consider gauge variables beside standard ones; making
use of the basis $\{ L_1 , ... , L_r \}$ we can take the $\a^m$ as
gauge variables. These index the gauge transformation, see
\eqref{eq:gauge} above. The function $\a : B \to \G$ is identified
with a section of a bundle $(\A_\G , \pi_\G , B)$, which is an
associated bundle to $P_G$ (the principal bundle of fiber $G$ over
$B$ defining the gauge action).

In this way, the phase bundle is augmented from $(M,\pi,B)$ to
$\^M = (M\oplus \A_\G  , \pi \oplus \pi_\G , B)$. Correspondingly
one would consider vector fields \beq \^X \ = \ \xi^i (x,u,\a)
\frac{\pa}{\pa x^i} \ + \ \phi^a (x,u,\a) \frac{\pa}{\pa u^a} \ +
\ A^m (x,u,\a) \frac{\pa}{\pa \a^m} \ ; \eeq we will actually
consider their evolutionary representatives, \beq X \ = \ Q^a
(x,u,\a;u_x,\a_x) \frac{\pa}{\pa u^a} \ + \ P^m (x,u,\a;u_x,\a_x)
\frac{\pa}{\pa \a^m} \eeq with components $$ Q^a = \phi^a - u^a_i
\xi^i \ , \ \ P^m = A^m - \a^m_i \xi^i \ . $$ Note these imply
that the $Q^a$ will not depend on $\a_x$, nor the $P^m$ on $u_x$.

If $X$ can be expressed as the gauge transformed of a vector field
in $M$, we say it is a {\it reframed} vector field. It is
immediate to see that such vector fields can be characterized as
those for which $Q$ can be written in the form \beq
\label{eq:reframed} Q^a (x,u,\a;u_x,\a_x) \ = \ [K^a_{\ b} (\a) ]
\ Q^b_0 (x,u,u_x)  \eeq with $K(\a) \in G$. One can choose $K(0) =
I$; in general we will have $K(\a ) = \exp [\a^m L_m]$. We then
say that $X$ is gauge equivalent to the vector field $Z = Q^a_0
(x,u,u_x) (\pa / \pa u^a)$.

When considering prolongations in $J \^M$, the prolongation
operation will also act on the components in the $\a^m$ direction.
Applying the (standard) prolongation formula on $X$ we will obtain
a vector field \beq\label{eq:augmprol} Y \ = \ \psi^a_J \,
\frac{\pa}{\pa u^a_J} \ + \ \chi^m_J \, \frac{\pa}{\pa \a^m_J} \ =
\ (\^D_J Q^a) \, \frac{\pa}{\pa u^a_J} \ + \ (\^D_J P^m) \,
\frac{\pa}{\pa \a^m_J} \eeq (the sum is also over all
multi-indices $J$ of suitable module $|J|$).

Here the notation $\^D_J$ recalls that total derivatives should be
computed by taking into account the gauge variables $\a^m (x)$ as
well: \beq \^D_i \ = \ \frac{\pa}{\pa x^i} \ + \ u^a_{J,i}
\frac{\pa}{\pa u^a_J} \ + \ \a^m_{J,i} \frac{\pa}{\pa \a^m_J} \ :=
\ D_i + Z_i \ . \eeq

If now we apply \eqref{eq:augmprol} on vector fields of the form
\eqref{eq:reframed}, we get -- for coefficients of first
derivative variables -- in explicit form \cite{Ggau1} \beq
\label{eq:prolgauge} \psi^a_i \ = \ (D_i Q^a) \ + \ (R_i)^a_{\ b}
\, Q^b \eeq where the matrices $R_i$ are defined as \beq R_i \ = \
[Z_i (K)] \ K^{-1} \ . \eeq These matrices satisfy \beq Z_i R_j \
- \ Z_j R_i \ + \ [R_i,R_j] \ = \ 0 \ ; \eeq this is nothing else
than the horizontal Maurer-Cartan equation\footnote{Note that as
the $K$ only depends on $\a$, actually $Z_i R_j = \^D_i R_j$;
hence this equation is equivalently written as $\^D_i R_j - \^D_j
R_i + [R_i,R_j] = 0$.} for the $R_i$. This also guarantees no
compatibility problem will arise when considering higher order
prolongations.

It is obvious that \eqref{eq:prolgauge} will reproduce
\eqref{eq:vproluno} if and only if $R_i = \La_i$. This relation
is, however, by itself meaningless as $R_i = R_i (\a,\a_x)$ and
$\La_i = \La_i (x,u,u_x)$. On the other hand, it makes sense if we
consider it {\it on a section of the gauge bundle} $(J^k \^M ,
\rho_k , J^k M)$ with fiber $\rho^{-1} (p) = \T^k \G_p$.

A section $\s_\ga$ of the gauge bundle is defined in local
coordinates by $$ \ga^m \ := \ \a^m \ - \ f^m (x,u) \ = \ 0 \ ; $$
its lift to first derivatives is hence given by $$ \ga^m_i \ := \
\a^m_i \ - \ \( \frac{\pa f^m}{\pa x^i} + u^a_i \frac{\pa f^m}{\pa
u^a} \) \ = \ \a^m_i \ - \ (D_i f^m) \ = \ 0 . $$

Therefore, the restriction of $R_i = \La_i$ to the section $\s$
makes perfect sense, and reads \beq \La_i \ = \ [\a^m_i]_\s \, L_m
\ = \ (D_i f^m ) \, L_m \ . \eeq

It should be noted that the restriction to a gauge section $\s$
makes sense provided this section is itself invariant under the
action of the vector field we are considering; this is the case
provided
$$ \[ P^m - (\pa_a f^m) Q^a \]_\s \ = \ 0 \ . $$
This can always be achieved by simply requiring $P^m = (\pa_a f^m)
Q^a$; in facts, our discussion constrained $Q$ to the form
\eqref{eq:reframed}, but did not set any constraint on $P$.
Finally, let us note that a section $\s$ of the gauge bundle $(J^k
\^M , \rho_k , J^k M)$ is by definition isomorphic to $J^k M$;
thus the vector field $X^{(k)}_\s$ (the restriction of $X$ to
$\s$) in $\s$ uniquely defines a vector field in $J^k M$.

In this way we have established a well-defined relation between
reframed (via a gauge transformation $K$) vector fields on $J^k
\^M$, restricted to a gauge section $\a$, and $\mu$-prolonged
vector fields on $J^k M$. The form $\mu$ is given by $\mu = \La_i
\d x^i$, and the $\La_i$ satisfy \beq \La_i \ = \ \[ K^{-1} \ Z_i
(K) \]_{\s^{(1)}} \ . \eeq

The reader is referred to \cite{Ggau1} for further details and
other results on this line.

\section{A gauge-theoretic approach to twisted symmetries}
\label{sec:gaugeB}

The discussion of the previous section (and of \cite{Ggau1}) was
based on considering the gauge variables as new dependent
variables; as seen above, this led to some inconsistency, which
could be cured only by restricting on a given section, i.e. {\it
de facto} forcing the gauge variables $\a^m$ to depend on $(x,u)$
in a given manner.

This situation is of course not satisfactory, and calls for a
fully coherent -- and fully gauge-theoretic -- treatment. This was
proposed in a recent paper \cite{Ggau2}, and we report here the
main lines of the construction proposed there. We also refer to
\cite{Ggau2} (see the appendix there) for a discussion of the
relation of this approach to that of \cite{Ggau1}, and also to the
approach to $\mu$-symmetries based on the formalism of coverings
\cite{CF2007}.

In the approach discussed in the previous section, gauge variables
were considered as new, auxiliary, dependent variables and treated
as such in complete parallel to standard dependent variables
(fields) up to the point where restriction to a section of the
gauge bundle was needed. This approach is quite unnatural to
anybody familiar with gauge theories, as gauge variables are
different than standard fields (matter fields in usual gauge
theories \cite{CCL,Nak,NaS}) and should be treated accordingly.
Moreover, the gauge variables control a change of reference frame,
which is the same for derivatives of any order -- in other words,
we do not need to consider prolongations of the gauge variables
beyond order one\footnote{The connection form is related to
derivatives of the gauge fields through the well known formula $A
= g^{-1} \d g$.}.

The key construction is still that of gauge bundles sketched in
section \ref{sec:gaugeA}; by this we mean both the basic gauge
bundle $(\rho, \^M , M)$ with fiber the gauge group $\Ga$ (i.e.
the set of maps $\ga : M \to G$), and the higher order gauge
bundles $(\rho_k, J^k \^M , J^k M)$ which also have fiber $\Ga$;
this is an important feature.

When one considers that $M$ is also a bundle $(\pi,M,B)$ and that
jet bundles $J^k M$ have several fiber structures, and in
particular can be seen as bundles both over $B$ -- in which case
we write $(\pi_k, J^k M, B)$ -- and over $M$ -- in which case we
write $(\s_k, J^k M, M)$ -- and moreover that the same
considerations hold for $J^k \^M$, the relations among all these
structures are embodied in the following ``star'' diagram
\cite{Ggau2}: \beq\label{diag:star} \matrix{J^k \^M &
\mapright{\rho_k} & J^k M \cr
 & & \cr
\mapdown{\^\s_k} & \matrix{\mapse{\^\pi_k} & & \mapsw{\pi_k} \cr &
B & \cr \mapne{\^\pi} & & \mapnw{\pi} \cr} & \mapdown{\s_k} \cr
 & & \cr
\^M & \mapright{\rho} & M \cr}  \eeq

As stressed above, we have (topologically) $J^k \^M = J^k M \times
\Ga $; that is, the jets only concern standard variables, and not
gauge ones. Hence, the prolongation operation does not involve the
latter.

Thus, the prolongation operation leading from $\^M$ to $J^k \^M$
should be based on the usual total derivative operators $D_i$, and
hence does {\it not} involve derivation with respect to the gauge
variables. On the other hand, a vector field in $\^M$ will have
components both in the $M$ and in the $G$ directions, the
prolongation operation should be applied {\it only } to the $M$
components.

As usual, it will be convenient to consider the vector bundle
associated to $J^k \^M$; in concrete terms, this means passing to
consider coordinates $\a^m$ in the Lie algebra $\G$ of $\Ga$,
similarly to what we made in the previous section\footnote{We can
also think of this operation as a restriction to a neighborhood of
a reference section in the gauge bundle, in which we can use Lie
algebra coordinates.}. For ease of notation we will keep the same
notation $J^k \^M$ for this.\footnote{A more careful discussion is
provided in \cite{Ggau2}, to which we refer for all missing
details.}

A vector field in $\^M$ will then be $\^X = \xi^i (x,u,\a ) (\pa /
\pa x^i) + \vphi^a (x,u,\a) (\pa / \pa u^a) + \ B^m (x,u,\a) (\pa
/ \pa \a^m)$. As usual in considerations involving vector fields
on jet bundles, it will be convenient to work with evolutionary
representatives \cite{Olv1,Ste}; we will consistently use these.
The evolutionary representative of $\^X$ is \beq\label{eq:Xv} X \
\equiv \ \^X_v \ = \ Q^a  \, {\pa \over \pa u^a} \ + \ P^m \, {\pa
\over \pa \a^m} \ , \eeq where $Q^a := \vphi^a - u^a_i \xi^i$,
$P^m := B^m$.

The coordinate expression of the prolongation $X^{(k)} \in \X (J^k
\^M)$ of $X$ is given by the (standard) prolongation formula; we
stress once again that no prolongation of $\a^m$ components
appears. For the evolutionary representative $Y := (X^{(k)})_v =
X_v^{(k)}$ we get, with $Q^a_J = D_J Q^a$, \beq\label{eq:Y1} Y \ =
\ Q^a_J \, {\pa \over \pa u^a_J} \ + \ P^m \, {\pa \over \pa \a^m}
\ . \eeq

We are specially interested in a particular class of vector
fields, i.e. those for which \beq\label{eq:sepglob} Q^a
(x,u,g;u_x) \ = \ [\Psi (g)]^a_{\ b} \ \Theta^b (x,u;u_x) \ . \eeq
These will be called, for obvious reasons, {\bf gauged vector
fields}.

Using local coordinates $(x,u,\a)$, eq. (\ref{eq:sepglob}) becomes
\beq\label{eq:sep} Q^a (x,u,\a;u_x) \ = \ [K(\a)]^a_{\ b} \
\Theta^b (x,u;u_x) \ . \eeq Here $K(\a)$ is the representation of
the group element $g (\a) = \exp (\a)$; i.e. if $G$ acts in $U$
via the representation $\Psi$, we have $K(\a ) = \Psi [ \exp
(\a)]$. Note that \eqref{eq:sepglob} and \eqref{eq:sep} do not
constrain in any way the components $P^m$ of the vector fields
along the $\a^m$ variables.

Keeping in mind that the total derivative operators do not act on
gauge variables, we obtain immediately that \beq Q^a_J \ = \ D_J
Q^a \ = \ [ K(\a )]^a_{\ b} \ D_J \Theta^b \ . \eeq This implies
that -- writing $\Theta_J = D_J \Theta$ -- the prolongation $Y$ of
the vector field $X$, see \eqref{eq:Y1}, is given by
\beq\label{eq:Y2} Y \ = \ [K(\a)]^a_{\ b} \, \Theta^b_J \, {\pa
\over \pa u^a_J} \ + \ P^m \, {\pa \over \pa \a^m}  \ . \eeq

Now, if $X_0 = \rho_* X$ and $Y_0 = \rho^{(k)}_* Y$ are the
projection of the vector fields $X$ and $Y$ to, respectively, $M$
and $J^k M$, we can state formally that {\it $Y_0$ is the
$\mu$-prolongation of $X_0$ for a suitable $\mu$.} In fact, we
have \beq X_0 = \rho_* X = [K(\a)]^a_{\ b} \, \Theta^b \, {\pa
\over \pa u^a} \ ; \ \ Y_0 = \rho^{(k)}_* Y = [K(\a)]^a_{\ b} \,
\Theta^b_J \, {\pa \over \pa u^a_J} \ . \eeq Thus $X_0$ and $Y_0$
are the gauge transformed -- via the same gauge transformation --
of vector fields $\=X_0$ and $\=Y_0$ such that $\=Y_0$ is the
ordinary prolongation of $\=X_0$.

Note this statement is only formal, as the $\a$ variables have no
meaning when we work in $M$ and $J^k M$; in order to make this
into a real theorem, we will need to ``fix the gauge'', as
discussed below (and similarly to what has been done in the
previous section). Note also the relation between $X_0$ and $Y_0$
depends substantially on the assumption that $X$ is a gauged
vector field.

Fixing the gauge means selecting a section $\ga$ of the gauge
bundle. We will denote by $\om^{(\ga)}$ the operator of
restriction from $\^M$ to $\^M_\ga$, and by $\rho^{(\ga)}$ the
restriction of the projection $\rho : \^M \to M$ to $\^M_\ga$. We
also denote by $\om^{(\ga)}_k : J^k \^M \to \^M_\ga^{(k)}$ and by
$\rho^{(\ga)}_k : \^M_\ga^{(k)} \to J^k M$ the lift of the maps
$\om^{(\ga)}$ and $\rho^{(\ga)}$ to maps between corresponding jet
spaces of order $k$.\footnote{Note that while $\rho$ is of course
not invertible, it follows from $M_\ga \simeq M$ that
$\rho^{(\ga)}$ is invertible, with $(\rho^{(\ga)})^{-1} = \ga$.
similarly, $\rho^{(\ga)}$ is invertible.}

We will summarize relations and maps between relevant fiber
bundles in the following diagram: \beq\label{diag:sectbundle}
 \matrix{ \^M & \mapright{\om^{(\ga)}} & \^M_\ga &
\mapright{\rho^{(\ga)}} & M \cr & & & & \cr \mapdown{j^{k} [\^M]}
& & \mapdown{j^{k} [\^M_\ga]} & & \mapdown{j^{k} [M]} \cr & & & &
\cr J^k \^M & \mapright{\om^{(\ga)}_k} & \^M_\ga^{(k)} &
\mapright{\rho^{(\ga)}_k} & J^k M \cr} \eeq

As anticipated, passing to consider $\^M_\ga$ rather than the full
$\^M$, and $\^M_\ga^{(k)}$ rather than the full $\^M^{(k)}$,
corresponds in physical terms to a {\it gauge fixing}.

Needless to say, restriction to $M_\ga$ makes sense only if this
manifold is invariant under the vector field we are considering;
it is easy to check that if $X$ be in the form \eqref{eq:Xv}, then
the submanifold $\^M_\ga \ss \^M$ identified by $\a^m = A^m (x,u)$
is invariant under $X_\ga$ if and only if \beq\label{eq:gainvar}
P^m_\ga \ = \ \( \pa A^m / \pa u^a \) \ Q^a_\ga  \ . \eeq As
remarked above, $P_m$ was not constrained by our previous
considerations, so it can be adjusted to the $M_\ga$, i.e. to the
gauge fixing $\ga$, we wish to consider. More precisely, one can
show \cite{Ggau2} that: given arbitrary smooth functions $Q^a
(x,u,\a)$, and an arbitrary section $\ga$ of the gauge bundle,
there is always a vector field $X_v$ of the form $X_v = Q^a \pa_a
+ P^m \pa_m$ such that $X_v$ leaves $\^M_\ga$ invariant.

It should also be noted that $X_\ga$ projects in turn to a
vertical vector field $W$ on $M$, $W = Q^a (x,u) (\pa / \pa u^a)$;
and conversely any such $W$ lifts to a vertical vector field
$W^\ga = X_\ga$ on $\^M_\ga$, $ W^\ga = Q^a (x,u) [ (\pa / \pa
u^a) + ( (\pa A^m / \pa u^a) (\pa / \pa \a^m)]$.

We can then consider gauging and prolongation of vector fields. In
particular, in studying twisted symmetries one is led to consider
if it is possible to find suitable operators ${\tt \^P}_\ga^{(k)}$
and ${\tt P}_\ga^{(k)}$ such that, for a given $X$ on $M$, we have
a commutative diagram \beq\label{diag:ZZ} \matrix{ X &
\mapright{\om^{(\ga)}_*} & X_\ga & \mapright{\rho^{(\ga)}_*} & W
\cr & & & & \cr \mapdown{\Pr{k} [\^M]} & & \mapdown{{\tt
\^P}_\ga^{(k)}} & & \mapdown{{\tt P}_\ga^{(k)}} \cr & & & & \cr
X^{(k)} & \mapright{(\om^{(\ga)}_k)_*} & X_\ga^{(k)} &
\mapright{(\rho^{(\ga)}_k)_*} & Y \cr} \eeq

We will just give the main results, referring again the reader to
\cite{Ggau2} for a complete discussion. We will introduce, in
order to state our result in a compact form, operators
$\tau^{(\ga)} := \rho^{(\ga)} \circ \om^{(\ga)}$, $\tau^{(\ga)} :
\^M \to M$; and correspondingly $\tau^{(\ga)}_k := \rho^{(\ga)}_k
\circ \om^{(\ga)}_k$, $\tau^{(\ga)}_k : \^M^{(k)} \to M^{(k)}$.

Then we have that: {\it The twisted prolongation operator ${\tt
P}_\ga^{(k)}$ is uniquely defined by the requirement that
$(\tau^{(\ga)}_k)_* \circ (\Pr{k} [\^M]) = {\tt P}_\ga^{(k)} \circ
\tau^{(\ga)}_*$. Moreover, ${\tt P}_\ga^{(k)}$ corresponds to the
$\mu$-prolongation operator of order $k$ with $\mu = [\d
\Psi(\ga)] \Psi(\ga^{-1})$. With the local coordinates $(x,u,\a)$,
this corresponds to $\mu = \La_i \d x^i$ where $\La_i = -
R_i^{(\ga)} = - (D_i K_\ga) K_\ga^{-1} = K_\ga (D_i K_\ga^{-1}
)$.}

A converse of the result above is as follows: {\it Let $Y$ be the
$k$-th $\mu$-prolongation of the evolutionary vector field $W$ on
$M$, with $\mu \in \La^1 (J^1 M , \psi (\G))$ given in coordinates
by $\mu = \La_i (x,u,u_x) \d x^i$. Then: {\tt (i)} there is a
section $\ga$ of the gauge bundle such that $ Y = {\tt P}_\ga
(W)$. {\tt (ii)} there is a vertical vector field $X$ in $\^M$
such that (\ref{diag:ZZ}) applies. {\tt (iii)} The matrix function
$K_\ga (x,u,u_x)$ satisfies $D_i K_\ga = - \La_i K_\ga$.}

To conclude this (long) section, let us briefly comment on the
geometrical meaning of the main results just given. We have shown
that the $\mu$-prolongation operator appears if we are insisting
in restricting our analysis to the phase bundle $M$ (or to the
sub-bundle $\^M_\ga \ss \^M$ seen as an image of $M$ under the
gauge map $\ga$ embedding it into $\^M$) rather than to the full
gauge bundle $\^M$. The fact we are considering projections of
vector fields in $\^M_\ga \ss \^M$ and ${\^M}_\ga^{(k)} \ss
\^M^{(k)}$ to vector fields in $M$ and $M^{(k)}$ makes that the
relation between basic vector fields and prolonged ones is not the
natural one, described by the prolongation operator, but is the
``twisted'' one described by the $\mu$-prolongation operator. See
also the discussion in the Appendix to \cite{Ggau2}.

\section{Twisted symmetries and variational problems}
\label{sec:variational}

As well known, symmetry analysis -- and symmetry reduction -- are
specially powerful in dealing with {\it variational problems}; the
key result relating standard symmetry and reduction is in this
framework the classical Noether theorem \cite{Noe}, see also
\cite{Kos,Olv1}.

It is thus not surprising that the application of twisted
symmetries to variational problems is specially fruitful and -- in
this author's opinion -- also specially fascinating.

Unfortunately, this paper is already way too long -- which also
accounts for the lack of detailed examples in it -- so that we
will just very sketchily mention the sources of the main results
obtained in this direction, leaving to the reader to look directly
the original papers\footnote{We hope to be able at a later time to
also review developments in this direction.}.

The first study in this direction was conducted by Muriel, Romero
and Olver \cite{MRO}; they considered variational problems defined
by a Lagrangian (of arbitrary order) in one dependent and one
independent variable, and studied both how $\la$-prolongations
allow to construct new methods for the reduction of Euler-Lagrange
equations and a version of Noether's theorem adapted to
$\la$-symmetries. In this case one focuses on ``variational $\la$
symmetries'' -- which are the twisted symmetries analogue of
standard variational symmetries \cite{Olv1} -- and obtains partial
conservation laws. With $L$ the Lagrangian, $X$ a vector field and
$Y$ its $\la$-prolongation, $X$ is a variational $\la$-symmetry of
$L$ if \beq Y (L) \ + \ L \ (D + \la ) \xi \ = \ (D + \la) B \eeq
for some function $B (x,u,u_x,...)$.

If the $n$-th order Lagrangian $L$ admits a variational
$\la$-symmetry, then there exists a Lagrangian $\^L$ of order
$(n-1)$ such that a $(2n-1)$-parameter family of solutions to the
variational problem described by $L$ can be found from the
solution to the variational problem described by $\^L$ by solving
an auxiliary first order equation. The result extends to
generalized variational $\la$-symmetries.

Moreover, denote by $E[L]$ the Euler-Lagrange equation
corresponding to $L$. If $L$ admits a variational $\la$-symmetry
$X$, and $Q$ is the characteristic of $X$, then there is some
$P(x,u,u_x,...)$ such that \beq Q \ E[L] \ = \ (D + \la ) \, P \ .
\eeq

It turns out \cite{MRO} that in this case $X$ is a $\la$-symmetry
for the equation $P[u] = 0$; the reduced equation obtained from
$P=0$ using the symmetry $X$ is, up to multipliers, the reduced
equation $E[\^L]$ for the variational problem (see above).

Finally, if one restricts on the solutions to the Euler-Lagrange
equation $E[L]=0$, then $(1/P) X$ is a variational symmetry of the
variational problem\footnote{When this happens, i.e. we need the
restriction to solutions of the Euler-Lagrange equation to have a
variational symmetry, one speaks of ``pseudo-variational
symmetries'' \cite{Olv1}.}. This allows to associate a partial
conservation law to such symmetries. We refer again the reader to
\cite{MRO} for details.

It should be stressed again that the Muriel-Romero-Olver approach
is not restricted to first order Lagrangians, and they consider
explicitly some higher order example. On the other hand, as
implied by consideration of $\la$-symmetries, they consider
Lagrangian problems with only one independent variable.

In her work on deformation of the Lie derivative \cite{Mor2007},
Morando considered variational $\la$- and $\mu$-symmetries. She
started from a geometrical characterization -- in terms of the
action of the deformed Lie derivative (see section
\ref{sec:gaugeA} above) on the Poincar\'e-Cartan form $\Theta$ --
of variational $\la$-symmetries, which reads \beq \L_Y^\mu (\Theta
) \ \in \ \E \ ; \eeq here $Y$ is the $\la$-prolongation of the
vector field $X$ on $M$, and $\E$ is the contact ideal (again, see
above).

This characterization is immediately generalized, and written
exactly in the same form, to $\mu$-symmetries and hence field
theory.

To a divergence variational $\la$-symmetry $X$ of a first order
regular Lagrangian $L$ is associated the ``$\la$-conservation
law'' \beq D_x \, (X \interno \Theta - R ) \ + \ \la \, (X
\interno \Theta - R) \ , \eeq where $R$ is a suitable smooth
function.

In the case of field theory, it is convenient to introduce a form
$$ \rho \ = \ R^i \Omega_i + \s \ , $$ where $\s \in \E$ and
$\Omega_i = \partial_i \interno \Omega$, with $\Omega = \d x^1 \w
... \w \d x^p$ the volume form on the base manifold $B$. In this
case one has a ``$\mu$-conservation law''; this reads \beq D_i \,
\interno \, \d^\mu (X \interno \Theta - \rho ) \ = \ 0 \eeq in
compact geometrical notation; in coordinates, and using freely the
standard notation introduced earlier on, it yields \beq \( D_i \ +
\ \Lambda_i \) \ \( \frac{\pa L}{\pa u^a_i} \, (\phi^a - u^a_k
\xi^k ) \, + \, \xi^i L \, - \, R^i \) \ = \ 0 \ . \eeq

In a different work \cite{CGnoe}, Cicogna and Gaeta also looked at
$\mu$-symmetries for variational problems; they also obtained that
to a $\mu$-symmetry of a Lagrangian is associated a deformed
conservation law as above.

On the other hand, they also looked at twisted symmetries of
variational problems from the point of view of gauged vector
fields. In this framework, it is natural to wonder how the
Euler-Lagrange equation themselves are transformed by a change of
reference frame. One thus obtains ``$\mu$-Euler-Lagrange
equations'', and this are exactly invariant under (the
$\mu$-prolongation of) a vector field which is a $\mu$-symmetry
for the underlying Lagrangian; correspondingly, a $\mu$-symmetry
for the Lagrangian yields an exact conservation law for the
associated $\mu$-Euler-Lagrange equations \cite{CGnoe}.

All these works were conducted in the Lagrangian framework; in a
very recent study, Cicogna considered the Hamiltonian counterpart
of these results. We will not discuss at all this aspect,
referring the reader to his paper \cite{Claham}.


\end{document}